\documentclass[twocolumn,letter]{jpsj2}

\def\gsim{\mathop {\vtop {\ialign {##\crcr 
$\hfil \displaystyle {>}\hfil $\crcr \noalign {\kern1pt \nointerlineskip
 } 
$\,\sim$ \crcr \noalign {\kern1pt}}}}\limits}
\def\lsim{\mathop {\vtop {\ialign {##\crcr 
$\hfil \displaystyle {<}\hfil $\crcr \noalign {\kern1pt \nointerlineskip
 } 
$\,\,\sim$ \crcr \noalign {\kern1pt}}}}\limits}

\def\runauthor{
Hiroyasu {\sc Matsuura} and Kazumasa {\sc Miyake}
}

\title{
Theory of Charge Kondo Effect on Pair Hopping Mechanism
}          

\author{Hiroyasu {\sc Matsuura}$^{1}$ and Kazumasa {\sc Miyake}$^{2}$ 
}
\inst
{
$^{1}$Department of Physics, University of Tokyo, 7-3-1 Hongo, Bunkyo, Tokyo 113-0033, Japan\\
$^{2}$Division of Materials Physics, Department of Materials Engineering Science \\
Graduate School of Engineering Science, Osaka University, Toyonaka, 
Osaka 560-8531, Japan\\
}

\recdate
{
\today
}

\abst
{
A new microscopic model of the charge Kondo effect is proposed, and is studied on the basis of a numerical renormalization group calculations.
It is shown that the charge Kondo effect is caused by a pair hopping interaction between the conduction band and the localized band.
It is found that the formation of a charge Kondo-Yosida singlet due to its mechanism corresponds to the appearance of valence skipping phenomenon, and that the charge Kondo-Yosida singlet and the spin Kondo-Yosida singlet state coexist for a tuned parameter set, because the charge and spin degrees of freedom are separated in this model.
Then, it is also found that the Sommerfeld coefficient is enhanced by the competition between these two singlet states.
}

\kword{Tl doped PbTe, Charge Kondo Effect, Pair Hopping, Negative-U Mechanism, Numerical renormalization group}
\begin{document}
\sloppy
\maketitle

It is well known that the formal valence of a Tl ion is 1+ or 3 +, 2+ being skipped, and that of Sn and Pb is 2+ and 4+, 3+ being skipped. 
The same phenomenon also appears for As, Bi, and so on. 
Namely, the formal number of $n$s electrons of these ions is either ($n$s$^0$) or ($n$s$^2$), while an ($n$s$^1$) state does not appear, where $n$ is the principal quantum number of an electronic state in centro-symmetric ions. 
This phenomenon is called ``valence skipping" or``inert pair effect".

To explain this phenomenon, many mechanisms have been proposed so far: An intra-atomic mechanism, an inter-atomic mechanism, a strong electron-phonon coupling mechanism, and so on~\cite{Anderson,Yoshida,Varma,Harrison,Hase,Hewson,Hotta}. 
It has been found that the phenomenon can be described by the attractive on-site interaction (i.e., negative-U model). 
Thus, on the basis of the negative-U model, the electronic states of compounds including the valence skipping elements have been studied, without asking the origin of the negative-U. 
 
Recently, a new type of Kondo effect has been observed in Tl doped PbTe~\cite{Matsushita1}.
Since Tl impurity has no magnetic moment, it has been suggested that the charge degeneracy of Tl ion is important for such a Kondo like effect to appear. 
Thus, this effect is called the charge Kondo effect.  
To explain the origin of the charge Kondo effect theoretically, the negative-U Anderson model~\cite{Taraphder} has been proposed, and the electronic state has been discussed in detail~\cite{Dzero,Costi}.

However, there are few discussions about the charge Kondo effect and the valence skipping phenomenon based on a microscopic model without assuming the negative-U. 
In addition, although novel physical properties are expected by the competition of the charge Kondo effect and the spin Kondo effect, it has not been discussed in detail. 

In this letter, we propose a microscopic mechanism based on electron correlations: It is shown that the charge Kondo effect is caused by a pair hopping interaction between a conduction band and a localized state, and is found that the charge Kondo-Yosida (KY) singlet and the spin KY singlet can coexist for a certain set of parameters in the microscopic model.

We propose an effective model Hamiltonian as follows:  
\begin{eqnarray}
\mathcal{H} = \mathcal{H}_0 + \mathcal{H}_{U}+ \mathcal{H}_{hyb},
\label{effmod}
\end{eqnarray}
where $\mathcal{H}_0 = \mathcal{H}_{c}+\mathcal{H}_{d}  + \mathcal{H}_{dc} + \mathcal{H}_{ph}$ with  
\begin{eqnarray}
\mathcal{H}_c& \equiv& \sum_{\bf{k}\sigma}(\epsilon_{\bf{k}} -\mu) c_{\bf{k}\sigma}^{\dag}c_{\bf{k}\sigma}, \label{cond} \\
\mathcal{H}_{d} &\equiv& (\Delta_{d} -\mu)\sum_{\sigma} n_{s\sigma}, \label{dlevel}\\
\mathcal{H}_{dc} &\equiv& U_{dc} \sum_{\bf{k} \bf{k}^{\prime} \sigma\sigma^{\prime}} c_{\bf{k}\sigma}^{\dag}c_{\bf{k}^{\prime} \sigma}n_{d\sigma^{\prime}},  \\
\mathcal{H}_{ph} &\equiv& J_{ph}\sum_{\bf{k} \bf{k}^{\prime}}\big[ d_{\uparrow}^{\dag}d_{\downarrow }^{\dag} c_{\bf{k}^{\prime}\downarrow} c_{\bf{k}\uparrow}+{\rm h.c.} \big], 
\end{eqnarray}
and
\begin{eqnarray}
\mathcal{H}_{U}&\equiv& U_{d}n_{d\uparrow}n_{d\downarrow}, \label{Uterm} \\
\mathcal{H}_{hyb}&\equiv& V_{dc}\sum_{\bf{k},\sigma} (c_{\bf{k},\sigma}^{\dag}d_{\sigma} +{\rm {h.c}}).  \label{hybterm}
\end{eqnarray}
Here, $d_{\sigma}$ and $c_{\bf{k}\sigma}$ are annihilation operators of an electron on the localized state and the conduction band with wave vector $\bf{k}$ and spin $\sigma$, and $n_{d\sigma}=d_{\sigma}^\dag d_{\sigma}$.
We denote the localized state as ``d" state.
$\epsilon_{\bf{k}}$ and $\mu$ are the dispersion of conduction band and the chemical potential, respectively.
$\Delta_{d}$, $U_{dc}$, and $J_{ph}$ are the one-body level of the localized state, the inter-orbital Coulomb interaction, and the pair hopping interaction between the conduction band and the localized state, respectively.
$U_{d}$ and $V_{dc}$ are the intra-orbital Coulomb interaction on the localized state and the hybridization between the localized state and the conduction band.

To clarify a role of the charge degree of freedom, we introduce pseudo-spin representations as follows:
\begin{eqnarray}
I^z_{d}   &\equiv& \frac{1}{2} (n_{d\uparrow} + n_{d\downarrow} -1), \label{axial1}\\
I^{+}_{d} &\equiv& d_{\uparrow}^{\dag}d_{\downarrow}^{\dag}, \label{axial2} \\
I^{-}_{d} &\equiv& d_{\downarrow}d_{\uparrow},  \label{axial3}
\end{eqnarray}
and
\begin{eqnarray}
I^z_{c}   &\equiv& \frac{1}{2} \sum_{{\bf k} {\bf k}^{\prime}}(c_{\bf{k}\uparrow}^{\dag}c_{\bf{k}^{\prime}\uparrow} + c_{\bf{k}\downarrow}^{\dag}c_{\bf{k}^{\prime}\downarrow} -\delta_{{\bf k}{\bf k}^\prime}), \label{axial4}\\
I^{+}_{c} &\equiv& \sum_{\bf{k}\bf{k}^{\prime}}c_{\bf{k}\uparrow}^{\dag}c_{\bf{k}^{\prime}\downarrow}^{\dag}, \label{axial5}\\
I^{-}_{c} &\equiv& \sum_{\bf{k}\bf{k}^{\prime}}c_{\bf{k}\downarrow}c_{\bf{k}^{\prime}\uparrow}. \label{axial6}
\end{eqnarray}
Hereafter, quantities defined by eqs. (\ref{axial1}) $\sim$ (\ref{axial6}) are called axial charges.
Then, the part $\mathcal{H}_0$ of model Hamiltonian (\ref{effmod}) is transformed as
\begin{eqnarray}
\mathcal{H}_0 \rightarrow \tilde{\mathcal{H}}_0 =\tilde{\mathcal{H}}_c + \tilde{\mathcal{H}}_{pot}  + \tilde{\mathcal{H}}_{dc} + \tilde{\mathcal{H}}_{ph} + \tilde{\mathcal{H}}_{d}, \label{effmod2}
\end{eqnarray}
where      
\begin{eqnarray}
\tilde{\mathcal{H}}_c&\equiv& \sum_{\bf{k}\sigma}(\epsilon_{\bf{k}} -\mu +U_{dc}) c_{\bf{k}\sigma}^{\dag}c_{\bf{k}\sigma},  \\
\tilde{\mathcal{H}}_{pot}&\equiv& \sum_{\bf{k}\bf{k}^{\prime} (\bf{k} \neq \bf{k}^{\prime})} U_{dc}c_{\bf{k}\sigma}^{\dag}c_{\bf{k}^{\prime}\sigma}, \\
\tilde{\mathcal{H}}_{dc} &\equiv& 4U_{dc}I^z_{d}I^z_{c}, \\
\tilde{\mathcal{H}}_{ph} &\equiv& J_{ph}(I^{+}_{c}I^{-}_{d}  + I^{-}_{c}I^{+}_{d}), \\
\tilde{\mathcal{H}}_{d} &\equiv& 2(\Delta_{d} -\mu +U_{dc})I_{d}^{z}. \label{Hd}
\end{eqnarray}
The effect of $\mathcal{H}_U$, (6), and $\mathcal{H}_{hyb}$, (7), will be discussed separately below. 
Here $\tilde{\mathcal{H}}_{pot}$, (16), is a potential scattering term. 
We find that this transformed model Hamiltonian, (\ref{effmod2}), corresponds to an anisotropic Kondo model with a potential scattering and a magnetic (polarized) field.
Therefore, it is expected that the Kondo effect is caused by the pair hopping interaction $J_{ph}$, if parameters are set adequately. 

We study the electronic states of Hamiltonian, (\ref{effmod2}), on the basis of the numerical renormalization group (NRG) method of Wilson~\cite{Wilson,KM,Bulla}.
The model Hamiltonians, (\ref{effmod2}), are transformed into the recursion form as follows:
\begin{eqnarray}
H_{N+1} &=& \Lambda^{1/2}H_{N} \nonumber \\ 
        && + \Lambda^{N/2}[\sum_{\sigma}\tau_{N}(f_{N\sigma}^{\dag}f_{N+1\sigma} + h.c.) \nonumber \\
        && + \sum_{\sigma}\epsilon_{N+1}f_{N+1\sigma}^{\dag}f_{N+1\sigma}], \label{NRG1}  
\end{eqnarray}
where $\Lambda$ is a scale factor (in this letter we set $\Lambda=3.0$), and the coefficients, $\tau_{N}$ and $\epsilon_{N}$, are estimated by the tridiagonal procedure~\cite{Bulla,Buxton}.    
$H_{N}$ is given by 
\begin{eqnarray}
H_{N} &=& \Lambda^{(N-1)/2}[\tilde{\mathcal{H}}_{dc}+ \tilde{\mathcal{H}}_{ph} + \tilde{\mathcal{H}}_{d} \nonumber  \\
     && +\sum_{\sigma}\sum_{n=0}^{N-1}\tau_{n}(f_{n\sigma}^{\dag}f_{n+1\sigma} + h.c.)  \nonumber \\
     && +\sum_{\sigma}\sum_{n=0}^{N}\epsilon_{n}f_{n\sigma}^{\dag}f_{n\sigma}].
\label{NRG2} 
\end{eqnarray}
The initial Hamiltonian $H_{0}$ is given by
\begin{eqnarray}
H_{0} &=& \Lambda^{-1/2}[\tilde{\mathcal{H}}_{dc}+ \tilde{\mathcal{H}}_{ph} + \tilde{\mathcal{H}}_{d} ].
\end{eqnarray}
Here we have discarded $\tilde{\mathcal{H}}_{pot}$, because $\tilde{\mathcal{H}}_{pot} \propto U_{dc}(1-\Lambda^{-1}) \rightarrow 0$ for $\Lambda \rightarrow 1$ according to the discussion in ref~\cite{KM}.
 
One can estimate the eigenvalues and eigenstates of $H_{N}$ by the repeated use of the recursion form with keeping 100 states in each iteration step.  
We also performed the NRG calculations with keeping 300 states, and verified that the result was the same as that keeping 100 states. 
 
In this letter, we show the entropy of the localized state and the fraction of the zero-, single-, and double occupancy on the localized state $\langle D_{Q_d} \rangle$, where $\langle D_{Q_d} \rangle$ takes $Q_d=0$ (for $d^{0}$ state), $Q_d=1$ (for $d^{1}$ state), and $Q_d=2$ (for $d^{2}$ state), respectively~\cite{Bulla}.

\begin{figure}[h]
\begin{center}
\rotatebox{0}{\includegraphics[angle=0,width=1\linewidth]{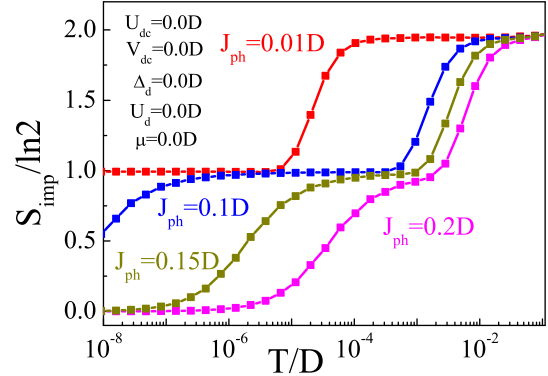}}
\caption{(Color Online) Temperature dependence of the entropy for $J_{ph}/D=0.01$, $0.1$, $0.15$, and $0.2$, respectively.
}
\label{Fig1}
\end{center}
\end{figure}

Figure \ref{Fig1} shows the temperature dependence of the entropy of the localized state for a series, $J_{ph} = 0.01D \sim 0.2D$ with $D$ being the half-bandwidth of the conduction band.
Other parameters are set as zero in order to see an essential character of the model.
Then, the model Hamiltonian, (\ref{effmod2}), corresponds to the anisotropic Kondo model.
For $J_{ph} =0.15D$, the entropy of the localized state is $S_{\rm{imp}} \sim k_B\rm{ln}4$ at $T/D \gsim 10^{-2}$, and $S_{\rm{imp}} \sim k_B\rm{ln}2$ at $10^{-5} \lsim T/D \lsim 10^{-3}$, and  $S_{\rm{imp}} \sim 0$ at $T/D \lsim 10^{-7}$.
We also find a similar behavior for $J_{ph} = 0.01D \sim 0.2D$.
In the region of $S_{\rm{imp}} = k_B\rm{ln}4$, the localized state is a free state where spin and charge degree of freedom are active.
On the other hand, in the region of $S_{\rm{imp}} = k_B\rm{ln}2$, there remains two degrees of freedom.
In the region of $S_{\rm{imp}} = 0$, the localized state is expected to be in the strong coupling singlet state.    

\begin{figure}[h]
\begin{center}
\rotatebox{0}{\includegraphics[angle=0,width=1\linewidth]{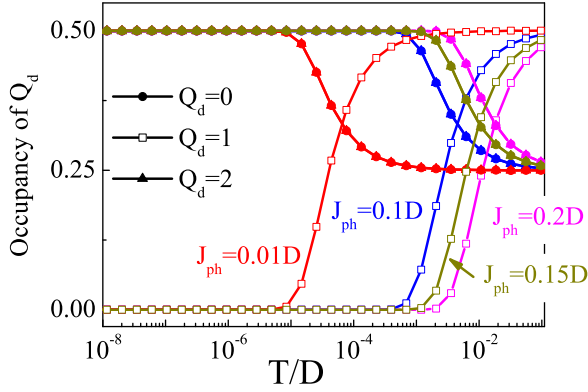}}
\caption{(Color Online) Temperature dependence on the zero-, single, and double occupancy for $J_{ph}/D=0.01$, $0.1$, $0.15$, and $0.2$, respectively.
$Q_{s}$ is the fraction of occupancy.
We set the same parameters as Fig. \ref{Fig1}.
}
\label{Fig2}
\end{center}
\end{figure}

To clarify the nature of the state with $S_{\rm{imp}}=k_B\rm{ln}2$, we estimate the fraction of the occupancy on the localized state.
Figure \ref{Fig2} shows its temperature dependence.
We use the same parameter set as Fig. \ref{Fig1}.
Here, we focus on the temperature dependence of entropy for $J_{ph} =0.15D$. At $T \gsim 10^{-2}D$ the fraction of occupancy is 0.25 for $Q_{d}=0$, 0.5 for $Q_{d}=1$, and 0.25 for $Q_{d}=2$, respectively. 
Then, we find that the electronic state of the localized state is the free state with 4 degrees of freedom.
At $T  \lsim 10^{-3}D$, one finds that the fraction of the occupancy of the d$^0$ state and the d$^2$ state are 0.5, while the fraction occupancy of the d$^1$ state is zero.
It means that the states of d$^0$ and d$^2$ states are degenerate. 
Thus, we conclude that $S_{\rm{imp}} = k_B{\rm ln}2$ stems from degenerate charge degrees of freedom (d$^0$ and d$^2$), implying that the valence skipping state is realized.
The same temperature dependence is observed also for $J_{ph}=0.01D \sim 0.2D$.

At low temperatures, we find that the energy level scheme of the fixed point is exactly the same as that of the strong coupling fixed point, although we do not show it in this letter.
Namely, we find that the ground state is the KY singlet state of the charge degree of freedom; the charge KY singlet state.   

When ${\mathcal{H}}_{U}$ and ${\mathcal{H}}_{hyb}$ are neglected, the model Hamiltonian of eq. (\ref{effmod}) or eq.(\ref{effmod2}) takes the same structure as the anisotropic Kondo model with a magnetic field.
By tuning the chemical potential ($\mu$) and the one-body potential ($\Delta_{d}$) so as to realize $\tilde{\mathcal{H}}_{d} =0$, the Hamiltonian (\ref{effmod2}) becomes the Kondo model without the magnetic field. 
Then, one finds that the ground state is the charge KY singlet by comparing with the above mentioned discussions.

When ${\mathcal{H}}_{U}$ and ${\mathcal{H}}_{hyb}$ are considered, the model Hamiltonian of eq. (\ref{effmod}) is not the same as the Kondo model.
However, it is shown that the Kondo effect naturally appears by tuning the parameters so as to realize a condition $E_{d^0} = E_{d^2}$, where $E_{d^i}$ is the energy of the d$^{i}$ state ($i=0,1,2$), estimated by the diagonalization of ${\mathcal{H}}_{U} + \tilde{\mathcal{H}}_{d}$ defined by eq. (\ref{Uterm})  and eq. (\ref{Hd}).
$E_{d^i}$ ($i=0,1,2$) is estimated as
\begin{eqnarray}
E_{d^0} &=& -\Delta_{d} + \mu -U_{dc} ,\\
E_{d^1} &=& 0 ,\\
E_{d^2} &=& U_d+\Delta_{d} - \mu +U_{dc}. 
\end{eqnarray}
Here, we analyze a simple particle-hole symmetric case corresponding to $-2\Delta_d=U_d$.
Then, $E_{d^0} =E_{d^2}$ is attained at $\mu =U_{dc}$.
We verified that qualitatively the same condition is attained, even if we relax the particle-hole symmetric condition.

\begin{figure}[h]
\begin{center}
\rotatebox{0}{\includegraphics[angle=0,width=1\linewidth]{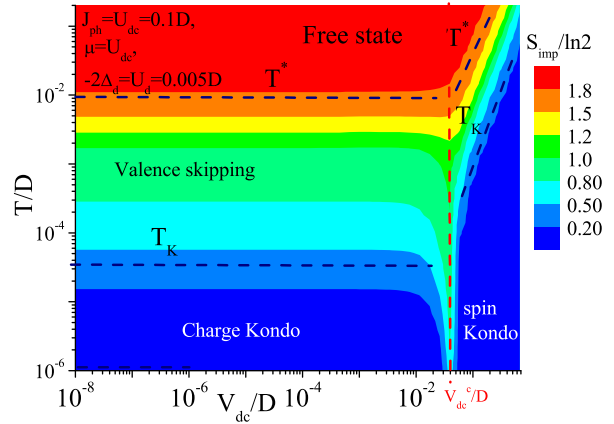}}
\caption{ (Color Online) Contour plot of the entropy of the localized state in $T/D-V_{dc}/D$ plane at $J_{ph}=U_{dc}=0.1D$, $\mu=U_{dc}$, and $-2\Delta=U_d=0.005D$.}
\label{Fig6}
\end{center}
\end{figure}

Figure \ref{Fig6} shows the contour plot of the entropy of the localized state in $T/D-V_{dc}/D$ plane. 
Here, we choose the parameter set as $J_{ph}=U_{dc}=0.1D$, $\mu=U_{dc}$, and $-2\Delta_d=U_d=0.005D$.

The temperature dependence of $S_{\rm{imp}}$ for $V_{dc}/D \lsim 10^{-2}$ is almost independent of $V_{dc}/D$; the electronic state changes from the free state to the valence skipping state at $T = T^{*} \simeq 10^{-2}D$, and the electronic state changes from the valence skipping state to the charge KY singlet state at $T = T_{K} \simeq 10^{-4}D$, as the temperature decreases, where $T^{*}$ is the temperature separating the free state and the valence skipping state (or the free spin state), and $T_{K}$ is the temperature separating the valence skipping state (or the free spin state) and the charge KY singlet state (the spin KY singlet state).
For $V_{dc}/D \gsim 4 \times 10^{-1}$, the ground state is the spin KY singlet state.
As $V_{dc}$ is increased, $T_{K}$ and $T^{*}$ increase.
For $V_{dc}/D =V_{dc}^{c}/D \simeq 3.659 \times 10^{-1}$, $T_{K}$ drastically decreases, and the entropy is $S_{\rm{imp}} =k_B \rm{ln}2$ down to technically zero temperature.
From the analysis of the energy flow diagram of NRG calculation, we find that the electronic state is the coexistence of the charge KY singlet and the spin KY singlet.
This coexistence is possible, because the charge and spin degree of freedoms are separated in this model.

\begin{figure}[h]
\begin{center}
\rotatebox{0}{\includegraphics[angle=0,width=1\linewidth]{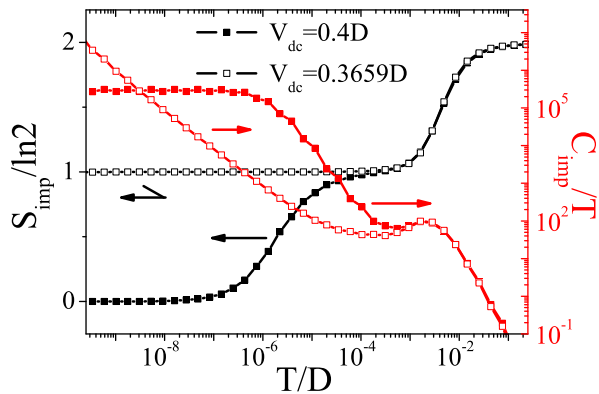}}
\caption{ (Color Online) Temperature dependence of entropy, $S_{\rm{imp}}$, (black line) and Sommerfeld coefficient $C_{\rm{imp}}/T$ (red line) for $V_{dc}=0.4D$ and $V_{dc}=0.3659D$. 
}
\label{SH}
\end{center}
\end{figure}

Figure \ref{SH} shows the temperature dependence of entropy, $S_{\rm{imp}}$, and the Sommerfeld coefficient, $C_{\rm{imp}}/T$, for $V_{dc}=0.4D$ and $V_{dc}=0.3659D$.
The temperature dependence of entropy for $V_{dc}=0.4D$ and $V_{dc}=0.3659D$ is consistent with the result in Fig. 3.
On the other hand, it is found that the Sommerfeld coefficient increases down to technically zero temperature for the seemingly critical value $V_{dc}^c \simeq V_{dc}=0.3659D$.
It is also found that the Sommerfeld coefficient increases in the region $10^{-6}D \lsim T \lsim 10^{-4}D$ for $V_{dc}=0.4D$, which is close to the critical value $V_{dc}^c$.
The origin of the enhancement is suspected to arise from the competition between the charge KY singlet and the spin KY singlet.
The details of the novel fixed point will be discussed elsewhere.

\begin{figure}[h]
\begin{center}
\rotatebox{0}{\includegraphics[angle=0,width=0.8\linewidth]{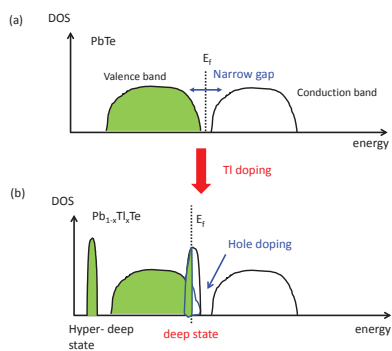}}
\caption{(Color Online) Schematic picture of the density of states in (a) PbTe and (b) Tl doped PbTe. 
The deep state is located at a lower energy level than the top of the valence band.
}
\label{Fig0}
\end{center}
\end{figure}

Finally, we discuss the possibility of the pair hopping mechanism in Tl doped PbTe.
The density of states (DOS) of PbTe is illustrated in Fig. \ref{Fig0}(a)~\cite{Heremans}. 
PbTe has a narrow gap where a direct gap is about $0.17$eV, and an indirect gap is about $0.025$ eV~\cite{Hase2}. 
When a Tl atom is doped in PbTe, the DOS of Pb$_{1-x}$Tl$_x$Te is illustrated as in Fig. \ref{Fig0}(b), where new two peaks appear: one peak near the top of the valence band is called a deep state, and the other peak much lower than the valence band is called a hyper-deep state~\cite{Hjalmarson}.
The energy level of the deep state is the characteristic property in Tl doped PbTe.
We find that the doping of Tl corresponds to the hole doping in the valence band.
Indeed, the direct hole doping into the valence band has been observed by the photoemission spectroscopy~\cite{Nakayama}.  

It is also known that the hyper-deep state and the deep state are almost localized states consisting of the s orbital of Tl and p orbitals of Te around the s orbital of Tl~\cite{Heremans,Ahmad}.
Thus, we can regard the deep state as the impurity orbital as in a d or f orbitals, while we neglect the hyper-deep state, because the energy level of the hyper-deep state is far below from the Fermi level.

In Tl doped PbTe, the pair hopping interaction is not the interaction between two atomic orbitals, but the interaction between the valence (conduction) band and the deep state.
The pair hopping interaction can be expressed by the linear combination of the atomic-Coulomb interactions between two p orbitals in Te.
It is expected that the amplitude of the pair hopping is of the order of the parameters as used in Fig. 3.
Thus, by tuning the chemical potential, it is expected that the charge Kondo effect is caused in a realistic parameter set as discussed in Fig. 3.

In conclusion, a new microscopic model of the charge Kondo effect has been proposed, and its model has been studied on the basis of the numerical renormalization group method.
It has been shown that the charge Kondo effect and the valence skipping phenomenon are caused by the pair hopping interaction between the conduction band and the localized band.
It has been also found that the charge and spin KY singlet states coexist in this effective model for the critical parameter set, and the Sommerfeld coefficient is enhanced by the competition between these singlet states.

\begin{acknowledgment}
One of us (H.M.) is grateful to C. M. Varma for fruitful discussions.
This work is supported by a Grant-in-Aid for Specially Promoted Research
(No. 20001004) from the Ministry of Education, Culture, Sports, Science and
Technology of Japan.
\end{acknowledgment}

\end{document}